\documentclass[aps,pra,twocolumn,amsmath,amssymb,showpacs,nofootinbib]{revtex4}

\newcommand{\bra}[1]{\langle#1|}
\newcommand{\ket}[1]{|#1\rangle}

\usepackage[dvips]{graphicx}
\usepackage{mathrsfs}

\begin{document}

\title{Modeling photo-detectors in quantum optics}

\author{Peter P. Rohde}
\email[]{rohde@physics.uq.edu.au}
\homepage{http://www.physics.uq.edu.au/people/rohde/}
\author{Timothy C. Ralph}
\affiliation{Centre for Quantum Computer Technology, Department of Physics\\ University of Queensland, Brisbane, QLD 4072, Australia}

\date{\today}

\begin{abstract}
Photo-detection plays a fundamental role in experimental quantum optics and is of particular importance in the emerging field of linear optics quantum computing. Present theoretical treatment of photo-detectors is highly idealized and fails to consider many important physical effects. We present a physically motivated model for photo-detectors which accommodates for the effects of finite resolution, bandwidth and efficiency, as well as dark-counts and dead-time. We apply our model to two simple well known applications, which illustrates the significance of these characteristics.
\end{abstract}

\pacs{03.67.Lx,42.50.-p}

\maketitle

\section{Introduction}
All experimental quantum optics schemes ultimately rely on photo-detection to perform measurements. The role of photo-detection is even more important in the emerging field of linear optics quantum computing (LOQC) \cite{bib:KLM01}, where in addition to providing measurement results, photo-detection is necessary for \emph{post-selection} or \emph{conditioning}, which is fundamental to all LOQC schemes.

Typically, theoretical treatment of photo-detectors is highly idealized in the sense that it restricts analysis to the photon-number degrees of freedom. In practice, photons have a modal structure far more elaborate than this, which includes, for example, spatial, spectral/temporal and polarization degrees of freedom. Photo-detectors themselves are sensitive to these additional degrees of freedom. Thus, accurate theoretical modeling of quantum optical systems necessitates a complete model for photo-detection which accommodates for effects in these additional degrees of freedom. This is especially the case when systems are subject to effects such as mode-mismatch \cite{bib:RohdeRalph05,bib:RohdePryde05,bib:RohdeRalph05b} which inherently occurs outside of the qubit space.

In this paper we present a physically motivated model for photo-detection which includes the effects of finite resolution, bandwidth and efficiency, as well as dark-counts and dead-time. We demonstrate our model by application to two simple example scenarios: conditional production of single photons from spontaneous down-conversion; and, Hong-Ou-Mandel interference \cite{bib:HOM87}.

We begin by reviewing present photo-detector models and discuss the modeling of finite detection efficiency, dead-time and dark-counts. We then discuss the effects of detector resolution and bandwidth, and present a model for photo-detection which accounts for these effects. Finally we demonstrate our model by applying it to the two example scenarios mentioned.

\section{Photon-number-resolving detectors} \label{sec:photon_num_res_det}
The ideal photo-detector is a device which responds to the photon-number of an incident state, and is able to distinguish between different number states. Such a detector can be modeled by the measurement projectors
\begin{equation} \label{eq:number_proj}
\hat\Pi_n=\ket{n}\bra{n}
\end{equation}
where $n$ is the measurement result, corresponding to the number of photons in the incident state. Clearly this set of projectors forms a legitimate POVM, since \mbox{$\sum_{i=0}^\infty \hat\Pi_n=\hat\openone$}.
Based on standard quantum measurement theory, the output state following measurement result $n$ is given by
\begin{equation}
\hat\rho_\mathrm{out}(n)=\hat\Pi_n\hat\rho\hat\Pi_n
\end{equation}
where normalization factors have been ignored (for simplicity we omit normalization factors throughout our discussion).

\section{Non-photon-number-resolving detectors}
Unfortunately, most presently available photo-detectors are unable to resolve the photon number of an incident state. Instead, they provide one of two possible measurement outcomes, or \emph{signatures}: a \emph{click} if one or more photons are incident on the detector; or, \emph{no click} if no photons are incident on the detector.
		
Such a detector can be modeled simply as an ideal photon-number-resolving detector, and tracing over all detection outcomes which result in the desired signature. The physical motivation behind such treatment is that at the quantum level the detector \emph{is} able to distinguish between different photon numbers, however the observer is unable to access this information. Thus, the output state associated with a detector \emph{click} can be expressed as
\begin{equation}
\hat\rho_\mathrm{out}(click)=\sum_{n=1}^\infty \hat\Pi_n\hat\rho\hat\Pi_n
\end{equation}
since $n\geq 1$ corresponds to the set of outcomes which trigger the desired signature. Similarly, a \emph{no click} detection event can be modeled as
\begin{equation}
\hat\rho_\mathrm{out}(no-click)=\hat\Pi_0\hat\rho\hat\Pi_0
\end{equation}

Many experimental quantum optics scenarios inherently require number-resolving photo-detection. This has motivated the examination of how non-number-resolving photo-detectors can be configured to approximate number-resolving detection. The basic principle of most such schemes is identical. An incident mode is spread across multiple, distinct output modes. In the limit of a large number of output modes, the probability of any given mode containing more than one photon becomes negligible. Thus, by measuring these modes independently using non-photon-number-resolving detectors, the sum of all the detection events approximates the number of photons in the incident state.

\begin{figure}[!htb]
\includegraphics[width=0.4\columnwidth]{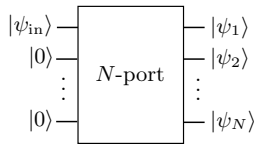}
\caption{Approximating a photon-number-resolving photo-detector using an $N$-port device. The $N$-port distributes the incident field evenly across the $N$ output modes, which are measured independently using non-number-resolving photo-detectors.} \label{fig:N_port}
\end{figure}

Most simply, such a scheme can be implemented using an $N$-port device \cite{bib:Kok01,bib:Paul96,bib:Bartlett02}, also referred to as a \emph{cascade network}. This is simply a beamsplitter network which equally distributes an incident field across $N$ output modes, as shown in Fig.~\ref{fig:N_port}. While very simple and elegant in principle, such a scheme requires $N$ beamsplitters, which becomes unwieldy for large $N$. A variation on this approach has been suggested which operates non-deterministically \cite{bib:Rohde05}, but which uses far fewer beamsplitters.

\begin{figure}[!htb]
\includegraphics[width=0.9\columnwidth]{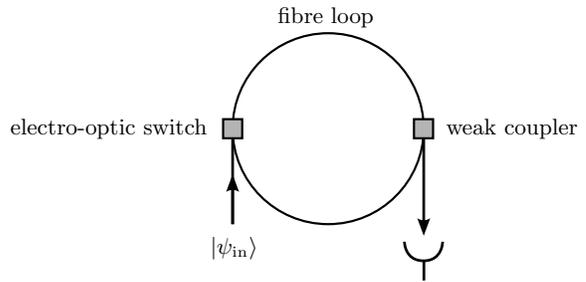}
\caption{Approximating a photon-number-resolving photo-detector using time-division multiplexing. The incident state is coupled into a low-loss fibre loop using an electro-optic switch. Following each round trip, each photon will have some probability of coupling out to the detector via the weak coupler.} \label{fig:TDM_detector}
\end{figure}

An alternate approach is to use \emph{time-division multiplexing} (TDM) \cite{bib:Achilles04,bib:Achilles03,bib:Banaszek03}, whereby the incident field is split into \emph{time-bins}. Each time-bin is then measured independently using a non-number-resolving photo-detector. Such a scheme can be implemented using a low-loss fiber loop, which inefficiently couples to a photo-detector, shown in Fig.~\ref{fig:TDM_detector}. After each round trip there is some low probability that a given photon will couple out to the detector. Otherwise it continues looping around the fiber. Thus, similar to the cascade approach, the photon arrival statistics are spread across multiple detection events to achieve low probability that multiple photons will trigger a single detection event.

A final approach is that employed by \emph{visible light photon counters} (VLPC's) \cite{bib:Kim99,bib:Takeuchi99,bib:Bartlett02}. Here an incident state is spatially dispersed and incident upon a detector array, which is able to respond independently to distinct \emph{spatial bins}. Again, the underlying principle is the same as before.

\section{Finite efficiency detectors}
Our treatment of photo-detectors thus far has been concerned with \emph{unit-efficiency detectors} -- ones which always respond to photons incident upon them. In practice, detectors always have finite efficiency. That is, even if a photon is incident upon a detector it need not necessarily trigger a detection event.
		
\begin{figure}[!htb]
\includegraphics[width=0.5\columnwidth]{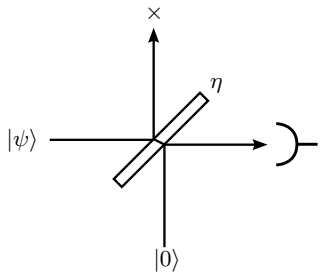}
\caption{Model for an inefficient photo-detector with quantum efficiency $\eta$. The mode transmitted through the beamsplitter is incident upon an ideal number-resolving photo-detector. The reflected mode is traced out.} \label{fig:inefficient_detector_model}
\end{figure}

Detector inefficiency can be modeled very simply by preceding a unit-efficiency detector with an $\eta$-transmissivity beamsplitter, as shown in Fig.~\ref{fig:inefficient_detector_model}. $\eta$ can be interpreted as the quantum efficiency of the detector or, equivalently, the probability that the detector will respond to any given photon. The transmitted modes are incident on the ideal photo-detector, while the reflected modes are traced out, thereby introducing mixing effects. The mixing arises as a result of ambiguity as to the true number of photons which trigger a given event. For example, in the presence of detector inefficiency an incident photon may still induce the zero photon signature.
	
Formally, the output state following the $n$ photon signature can be expressed
\begin{equation} \label{eq:fin_eff_det}
\hat\rho_\mathrm{out}=\hat\Pi_n\mathrm{tr}_\mathrm{ref}\left[\hat{U}_\mathrm{BS}\hat\rho\hat{U}_\mathrm{BS}^\dag\right]\hat\Pi_n
\end{equation}
where $\hat{U}_\mathrm{BS}$ is the beamsplitter operation with transmissivity $\eta$, and we trace over the reflected mode.

\section{Dead-time}
An effect related in nature to detector efficiency is that of \emph{dead-time}. When a photo-detector \emph{clicks} there is typically a period after this during which the detector will be unresponsive. Such a phenomenon can be modeled using a variation of the inefficient detector model presented previously. Instead of modeling the detector using constant efficiency $\eta$, we introduce a time-dependent efficiency function $\eta(t)$. This function has the property that before a \emph{click} it simply assumes the value of the detector's efficiency. However, immediately following a \emph{click} it drops to zero for a period equal to the dead-time, after which it regenerates to its former value. The exact form of the function will depend on the physical processes causing dead-time. However, as a simplifying approximation one could employ an inverted top-hat function of the form
\begin{equation}
\eta(t)=\left\{ \begin{array}{ll}
	\eta_\mathrm{eff}, & t<t_\mathrm{click}\\
	0, & t_\mathrm{click}\leq t<t+\tau_\mathrm{dead}\\
	\eta_\mathrm{eff}, & t\geq t+\tau_\mathrm{dead}
	\end{array} \right.
\end{equation}
where $\eta_\mathrm{eff}$ is detector efficiency, $t_\mathrm{click}$ is the time at which a \emph{click} event occurs, and $\tau_\mathrm{dead}$ is the dead-time.

\section{Dark-counts}
In the models presented so far we have assumed the complete absence of \emph{dark-counts}, whereby a detector falsely responds to non-existent or unwanted photons. This will introduce mixing effects in a manner similar to detector inefficiency. That is, there will be ambiguity as to the meaning of a given detection signature. Fortunately the effects of dark-counts can be kept extremely low using present day experimental techniques and can often be ignored. However, for completeness, we now consider how the effects of dark-counts can be modeled.

\begin{figure}[!htb]
\includegraphics[width=0.5\columnwidth]{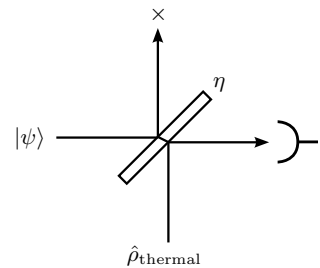}
\caption{Jointly modeling dark-counts and detector inefficiency using a beamsplitter and ideal photo-detector. The beamsplitter mixes the incident state with a thermal source, which is then incident on the photo-detector.} \label{fig:dark_count_model}
\end{figure}

We model dark-counts by assuming that their physical origin is stray photons entering from the environment. We assume the unwanted photons to be a thermal source, which couple into the detector. This can be modeled with a simple modification of our previous model for an inefficient detector. Instead of assuming a vacuum state to be incident on the unused beamsplitter port, we now assume a thermal state of the form
\begin{equation} \label{eq:thermal_source}
\hat\rho_\mathrm{thermal}=\frac{1}{\mathrm{cosh}(r)}\sum_{n=0}^\infty \mathrm{tanh}(r)^n\ket{n}\bra{n}
\end{equation}
to be incident. This models a thermal source of temperature \mbox{$T=\mathrm{e}^{2r}-1$} and average photon number \mbox{$\langle\hat{N}\rangle=\mathrm{e}^{2r}\mathrm{sinh}(r)$}. For any given $\eta$, an appropriate choice of $r$ can model an arbitrary dark-count rate. This model is illustrated in Fig.~\ref{fig:dark_count_model}.

With a thermal source of the form shown in Eq.~\ref{eq:thermal_source}, the probability of $k$ dark-counts occurring (\emph{i.e.} of $k$ photons from the thermal source being reflected by the beamsplitter) can be expressed
\begin{equation}
p_\mathrm{dc}(k)=\frac{1}{\mathrm{cosh}(r)}\sum_{n=k}^\infty \binom{n}{k}\mathrm{tanh}(r)^n\eta^{n-k}(1-\eta)^k
\end{equation}
It follows that the input state following detection can be expressed
\begin{equation}
\hat\rho_\mathrm{out}(n)=\hat\Pi_n\hat\rho'\hat\Pi_n+\sum_{i=0}^{n-1}p_\mathrm{dc}(n-i)\hat\Pi_i\hat\rho'\hat\Pi_i
\end{equation}
where $\hat\rho'=\mathrm{tr}_\mathrm{ref}\left[\hat{U}_\mathrm{BS}\hat\rho\hat{U}_\mathrm{BS}^\dag\right]$. In other words, given the $n$ photon signature, in addition to the desired term, the mixture will contain additional terms $i$ for all photon numbers less than $n$, where the probability associated with each of these terms is the probability of $n-i$ dark-counts occurring.

\section{Infinite resolution detectors}
The various detector models discussed so far make varying assumptions about their response to photon-number. We now go on to consider their behavior in non-photon-number degrees of freedom. We specifically consider detector spectral/temporal effects. However the discussion and models presented can be trivially modified to encompass other degrees of freedom (\emph{e.g.} spatial and polarization).
		
If we assume a detector has a delta function spectral response, \emph{i.e.} the detector responds only to photons of infinitesimal frequency $\omega$, the measurement projector for a single photon detection event can be expressed
\begin{equation}
\hat\Pi_1(\omega)=\ket{\omega}\bra{\omega}
\end{equation}
where $\ket\omega=\hat{a}^\dag(\omega)\ket{0}$ is the single photon state at frequency $\omega$, and the subscript `1' indicates that this is a single-photon projector. Such a detector, in addition to telling us that we have exactly one photon, gives us perfect information about its frequency, \emph{i.e.} with infinite spectral resolution.
		
\section{Finite resolution detectors}
In practice, no detector, even in principle, can have infinite spectral resolution. There will necessarily be some uncertainty in the frequency of an incident photon for a given detection signature. We now consider how we might generalize our model to accommodate for this.

We begin by assuming a detector which is the closest analogue of the $\hat\Pi_1$ detector, discussed in Sec.~\ref{sec:photon_num_res_det}. That is, a detector which \emph{clicks} if exactly one photon is incident upon it, and does not provide \emph{any} information about its frequency. Such a detector can be modeled by the projector
\begin{equation} \label{eq:zero_res_det}
\hat\Pi_1=\int_{-\infty}^{\infty}\hat\Pi_1(\omega)\,\mathrm{d}\omega=\int_{-\infty}^{\infty}\ket{\omega}\bra{\omega}\,\mathrm{d}\omega
\end{equation}
It can be seen upon inspection that this projector responds to the single photon state, while performing the identity operation in frequency space. It is not able to distinguish between different frequency components of a single photon state and consequently does not collapse the spectral wave-function. Specifically,
\begin{equation}
\hat\Pi_1\int_{-\infty}^{\infty}\psi(\omega)\ket{\omega}\,\mathrm{d}\omega=\int_{-\infty}^{\infty}\psi(\omega)\ket{\omega}\,\mathrm{d}\omega
\end{equation}
		
This idea can be logically generalized to the $n$ photon case as follows,
\begin{equation}
\hat\Pi_n=\int_{-\infty}^{\infty}\!\dots\!\int_{-\infty}^{\infty}\hat\Pi_1(\omega_1)\otimes\dots\otimes\hat\Pi_1(\omega_n)\,\mathrm{d}\omega_1\dots\mathrm{d}\omega_n
\end{equation}
		
We now have models for two unrealistic types of photo-detectors, with infinite and zero resolution respectively. We are now in a position to consider the more realistic case of finite resolution photo-detection (\emph{i.e.} where the resolution is greater than zero, but less than infinity). Such detectors will only respond to a certain range of frequencies, and with varying efficiency. This can be modeled by adapting our technique for modeling detector inefficiency. Specifically, we precede a zero-resolution detector of the form shown in Eq.~\ref{eq:zero_res_det}, with a beamsplitter with frequency-dependent transmissivity $\eta(\omega)$. In principle $\eta(\omega)$ can assume an arbitrary form. We might expect that in reality the form would be approximately Gaussian. The exact form will be defined by the physical processes which constitute the detection process (for example, the absorption of a photon by an atom, or the reflective properties of the front window).
		
The state following such a projection can be expressed in exactly the same form as Eq.~\ref{eq:fin_eff_det}, where $\hat{U}_\mathrm{BS}$ is now the frequency dependent beamsplitter operation.
		
In general, such a model is analytically complicated to work with due to the introduction of additional modes by the beamsplitter. The expression can be simplified significantly by introducing the approximation that $\eta(\omega)$ is a top-hat function. In this case the beamsplitter is always either completely transmissive or completely reflective, mitigating the necessity to introduce additional modes and perform analytically complicated trace-out operations. It can easily be shown that the measurement projector is then formally equivalent to
\begin{equation}
\hat\Pi_{1,\omega_0}=\int_{\omega_0-\delta}^{\omega_0+\delta}\hat\Pi(\omega)\,\mathrm{d}\omega
\end{equation}
where $\omega_0$ is the centre frequency of the top-hat and $\delta$ its half-width. We refer to $\delta$ as the \emph{resolution} of the detection event. We will assume that a projection of this form corresponds to a classical readout of frequency $\omega_0$. Thus, this detector, in addition to providing information about photon number, provides an additional measurement parameter, the frequency of the detection event, $\omega_0$, with finite resolution $2\delta$. Such a detector cannot, in principle, discriminate between different frequencies in the range $\omega_0-\delta$ to $\omega_0+\delta$.	 The output state associated with the single photon detection signature $\omega_0$ can be expressed	
\begin{equation} \label{eq:inf_res_det}
\hat\rho_\mathrm{out}(\omega_0)=\hat\Pi_{1,\omega_0}\hat\rho\hat\Pi_{1,\omega_0}
\end{equation}

The model for a finite resolution detector can be easily generalized to the multi-photon case,
\begin{eqnarray}
&&\hat\Pi_{n,\vec\omega_0}=\nonumber\\
&&\int_{\omega_1-\delta}^{\omega_1+\delta}\!\dots\!\int_{\omega_n-\delta}^{\omega_n+\delta}\hat\Pi_1(\omega_1)\otimes\dots\otimes\hat\Pi_1(\omega_n)\,\mathrm{d}\omega_1\dots\mathrm{d}\omega_n\nonumber\\
\end{eqnarray}
where $\vec\omega_0=\{\omega_1,\dots,\omega_n\}$ is the set of measurement results for each of the $n$ photons and we have again made the top-hat assumption.

\section{Finite resolution detectors with imperfect information}
We have discussed the modeling of photo-detectors with finite spectral resolution, where a detection signature $\omega_0$ is associated with some degree of uncertainty in the actual frequency of the incident photon. While this may accurately describe the operation of actual photo-detectors at the microscopic level, it typically does not accurately describe macroscopic operation. Specifically, although a detector may produce the classical information $\omega_0$ associated with a given detection event, it may be inaccessible to the observer. Instead the observer may simply observe the sum of the signals produced by the various microscopic detection events. For example, in \emph{time-integrated detection} the observer sees a signal representing the sum of detection events in a given time window, without gaining any information about the arrival times of the photons within that window. Such effects clearly introduce mixture into the state of systems. We now consider a more generalized photo-detector model which includes such effects.

\begin{figure}[!htb]
\includegraphics[width=\columnwidth]{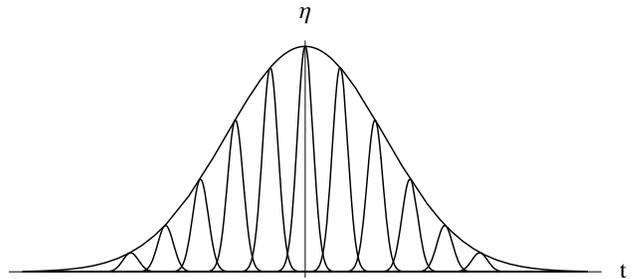}
\caption{Conceptualization of the detector model. The smaller inner curves represent the response functions of the different quantum level detection events, which are classically distinguishable from one another. The outer envelope represents the macroscopic response of the detector. Knowledge of which quantum level detection event occurred is not available to the observer. Thus, the state following photo-detection corresponds to the incoherent sum of all microscopic detection events within the macroscopic envelope.} \label{fig:detector_response}
\end{figure}

For simplicity, we discuss the single-photon case. However, as previously, the model can be logically generalized to the multi-photon case. The output state following detection of a single photon, where spectral information is not accessible by the observer, will be an incoherent superposition of the form
\begin{equation} \label{eq:gen_fin_res_det_single}
\hat\rho_\mathrm{out}(click)=\int_{-\infty}^{\infty}\hat\Pi_1\mathrm{tr}_\mathrm{ref}\left[\hat{U}_\mathrm{BS}(\omega_0)\hat\rho\hat{U}_\mathrm{BS}^\dag(\omega_0)\right]\hat\Pi_1\,\mathrm{d}\omega_0
\end{equation}
The beamsplitter operation $\hat{U}_\mathrm{BS}$ is now a function of both frequency and the specific measurement result $\omega_0$. Its transmissivity function can be expressed in the form $\eta(\omega_0,\omega)$. This captures the fact that different quantum level detection events (\emph{i.e.} different $\omega_0$) will occur with differing efficiencies and spectral responses.  Probability laws enforce the constraint
\begin{equation}
\int_{-\infty}^{\infty}\eta(\omega_0,\omega)\,\mathrm{d}\omega_0=\eta_\mathrm{eff}(\omega)\leq1
\end{equation}
where $\eta_\mathrm{eff}(\omega)$ is the effective efficiency of the detector at frequency $\omega$. In other words, for an incident photon of frequency $\omega$, the sum of the probabilities of all available quantum level detection events occurring can be interpreted as the detector's net efficiency at that frequency, which cannot exceed unity. Fig.~\ref{fig:detector_response} illustrates a conceptualization of this model.

If we again make the simplifying assumption that the detector has a top-hat response (at both the microscopic and macroscopic levels), the state following such a measurement can be expressed
\begin{equation} \label{eq:gen_photo_det_model}
\hat\rho_\mathrm{out}(click)=\int_{\Omega_0-\Delta}^{\Omega_0+\Delta}\hat\Pi_{1,\omega_0}\hat\rho\hat\Pi_{1,\omega_0}\mathrm{d}\omega_0
\end{equation}
where $\Omega_0$ is the macroscopic centre frequency of the detector, and $\Delta$ the macroscopic spectral half-width which stipulates the range of frequencies the detector responds to. We refer to $\Delta$ as the \emph{bandwidth} of the detector. The parameters $\omega_0$ and $\delta$ correspond to particular quantum level detection events. In Fig.~\ref{fig:detector_response}, $\Delta$ can be interpreted as the width of the outer envelope and $\delta$ as the width of the smaller inner curves.

\section{Application I: Conditional production of single photons from spontaneous down-conversion}
One of the simplest applications of photo-detection where spatio-temporal effects are of significance is in the conditional production of single photons from entangled pairs produced through spontaneous parametric down-conversion sources. Such sources probabilistically emit pairs of photons into distinct spatial modes which are entangled in their spatio-temporal degrees of freedom. Detection of a photon in one spatial mode implies the presence of a photon in the other mode with high probability. The production of single photons using this technique is widely employed in present-day quantum optics experiments and has been extensively studied \cite{bib:Rubin94,bib:Ou97,bib:Grice01,bib:URen03,bib:URen05,bib:URenMukamel03,bib:OuLu99,bib:Aiello98,bib:Aichele02,bib:Bellini03}.
		
We consider a general source, which emits photon pairs of the form
\begin{equation}
\ket\psi=\int_{-\infty}^{\infty}\int_{-\infty}^{\infty}\psi(\omega_1,\omega_2)\hat{a}^\dag(\omega_1)\hat{b}^\dag(\omega_2)\,\mathrm{d}\omega_1\,\mathrm{d}\omega_2\,\ket{0}
\end{equation}
where $\psi(\omega_1,\omega_2)$ is the joint spectral wave-function of the two-photon state.
		
Upon conditioning on the detection of a single photon in mode $a$, using the photo-detector model of Eq.~\ref{eq:gen_photo_det_model} (\emph{i.e.} when employing the top-hat simplifications), the reduced state of the other (\emph{signal}) mode is characterized by
\begin{eqnarray} \label{eq:gen_cond_state}
&&\hat\rho_\mathrm{signal}=\int_{\Omega_0-\Delta}^{\Omega_0+\Delta}\left(\int_{-\infty}^{\infty}\int_{\omega_0-\delta}^{\omega_0+\delta}\psi(\omega_1,\omega_2)\ket{\omega_2}\,\mathrm{d}\omega_1\,\mathrm{d}\omega_2\right)\nonumber\\
&&\times\left(\int_{-\infty}^{\infty}\int_{\omega_0-\delta}^{\omega_0+\delta}\psi^*(\omega_1,\omega_2)\bra{\omega_2}\,\mathrm{d}\omega_1\,\mathrm{d}\omega_2\right)\mathrm{d}\omega_0\nonumber\\
\end{eqnarray}
where we have traced out the detected modes so as to only consider the final state of the undetected mode.
		
It is illustrative to consider the nature of this state in some limiting cases. When $\delta\ll\Delta$ and $\delta\ll\sigma(\psi)$ (\emph{i.e.} in the limit where the uncertainty in the quantum level detection events is much less than both the detector and source bandwidths), the output state approaches the form
\begin{eqnarray}
\hat\rho_\mathrm{signal}\approx\int_{\Omega_0-\Delta}^{\Omega_0+\Delta}\left(\int_{-\infty}^{\infty}\psi(\omega_0,\omega_2)\ket{\omega_2}\,\mathrm{d}\omega_2\right)\nonumber\\
\times\left(\int_{-\infty}^{\infty}\psi^*(\omega_0,\omega_2)\bra{\omega_2}\,\mathrm{d}\omega_2\right)\mathrm{d}\omega_0
\end{eqnarray}
The form of this state is equivalent to that reported in Ref.~\cite{bib:URen05} for the conditional production of single photons from entangled pairs using time-integrated detection. This is expected since time-integrated detection corresponds to having detectors with delta function response, but where this information is not accessible to the observer.
		
The nature of this state varies enormously depending on the form of $\psi(\omega_1,\omega_2)$. Perhaps the most interesting cases are in the limits where the incident two-photon term is either maximally entangled or completely separable. Let us first consider the former case. Here all the off-diagonal terms of $\psi(\omega_1,\omega_2)$ are zero and the state reduces to
\begin{equation}
\hat\rho_\mathrm{signal}\approx\int_{\Omega_0-\Delta}^{\Omega_0+\Delta}\left|\psi(\omega_0,\omega_0)\right|^2\ket{\omega_0}\bra{\omega_0}\,\mathrm{d}\omega_0
\end{equation}
Despite having well defined photon number, this state is completely spectrally mixed. This comes about because \emph{knowledge} of which spectral component was detected in the other photon (but which is hidden from the observer) manifests itself in the conditionally prepared photon via the spectral entanglement between the two. The spectral distribution of the conditionally produced photon is preserved, but truncated according to the macroscopic bounds of the detector's response. Now let us consider the opposing limit of spectrally separable photons. Now the joint spectral wave-function can be factorized, $\psi(\omega_1,\omega_2)=\psi_\mathrm{A}(\omega_1)\psi_\mathrm{B}(\omega_2)$, and the conditionally prepared state expressed
\begin{eqnarray}
&&\hat\rho_\mathrm{signal}\approx\int_{\Omega_0-\delta}^{\Omega_0+\delta}\left|\psi_\mathrm{A}(\omega_0)\right|^2\,\mathrm{d}\omega_0\nonumber\\
&&\times\left(\int_{-\infty}^{\infty}\psi_\mathrm{B}(\omega_2)\ket{\omega_2}\,\mathrm{d}\omega_2\right)\left(\int_{-\infty}^{\infty}\psi_\mathrm{B}^*(\omega_2)\bra{\omega_2}\,\mathrm{d}\omega_2\right)\nonumber\\
\end{eqnarray}
Unlike the former case, this state is spectrally pure. This is equivalent to the observation made in Ref.~\cite{bib:URen05} that to conditionally prepare pure single photon states, the incident entangled pair should be separable in the spatio-temporal degrees of freedom.

We now consider the opposing detection limit where $\Delta\ll\delta$. Physically, this corresponds to a detector in which only a single microscopic level detection event may occur. In other words, the detector does not hide any information from the observer. Now the state can be approximated as
\begin{eqnarray}
&&\hat\rho_\mathrm{signal}\approx\left(\int_{-\infty}^{\infty}\int_{\Omega_0-\delta}^{\Omega_0+\delta}\psi(\omega_1,\omega_2)\ket{\omega_2}\,\mathrm{d}\omega_1\,\mathrm{d}\omega_2\right)\nonumber\\
&&\times\left(\int_{-\infty}^{\infty}\int_{\Omega_0-\delta}^{\Omega_0+\delta}\psi^*(\omega_1,\omega_2)\bra{\omega_2}\,\mathrm{d}\omega_1\,\mathrm{d}\omega_2\right)
\end{eqnarray}
In stark contrast to the previous example, this state is spectrally pure, regardless of the form of $\psi(\omega_1,\omega_2)$. The original spectral distribution is once again preserved, but, depending on the degree of entanglement between the two photons, may be truncated according to the bounds of the microscopic detection event.

Based on these example scenarios we may quickly establish the following criteria for the conditional production of single photons from entangled pairs: the spatio-temporal entanglement in the incident two-photon state should be minimized, and $\delta$ should be as large as possible with respect to $\Delta$. It is for the later reason that experimental realizations of such schemes typically employ very narrowband filtering and irising, since this effectively reduces $\Delta$ while leaving $\delta$ unchanged.

\section{Application II: Hong-Ou-Mandel interference}
\begin{figure}[!htb]
\includegraphics[width=0.8\columnwidth]{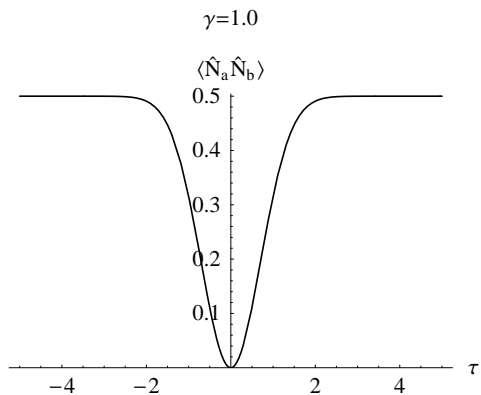}
\caption{Ideal HOM dip where no mode-mismatch is present. $\langle\hat{N}_\mathrm{a}\hat{N}_\mathrm{b}\rangle$ is the coincidence probability and $\tau$ is the magnitude of the temporal delay introduced into one of the photons (in units of the photons' temporal bandwidth).} \label{fig:HOM_dip_ideal}
\end{figure}

We now consider application of our photo-detector model to Hong-Ou-Mandel (HOM) \cite{bib:HOM87} interference. Here two photons are incident upon a 50/50 beamsplitter. Ideally, photon bunching occurs, resulting in the complete suppression of terms where a single photon is present at each beamsplitter output. Specifically,
\begin{equation}
\hat{U}_\mathrm{BS}\ket{1}_\mathrm{a}\ket{1}_\mathrm{b}\to\frac{1}{\sqrt{2}}\left(\ket{2}_\mathrm{a}\ket{0}_\mathrm{b}-\ket{0}_\mathrm{a}\ket{2}_\mathrm{b}\right)
\end{equation}
where $\hat{U}_\mathrm{BS}$ is the beamsplitter operation with $\eta=0.5$. Photon distinguishability undermines this bunching effect, resulting in \emph{coincidence} events, where a single photon is detected at each output. Thus, when the photons are perfectly matched on the beamsplitter, the coincidence rate is zero. However as we introduce photon distinguishability, typically implemented via a temporal delay in one photon, the coincidence probability asymptotically approaches $1/2$, the classical limit. This behavior is illustrated in Fig.~\ref{fig:HOM_dip_ideal} -- the well known \emph{HOM dip}.

The operation of the scheme is typically quantified in terms of the \emph{HOM visibility}, defined as
\begin{equation}
V=\frac{\langle\hat{N}_\mathrm{a}\hat{N}_\mathrm{b}\rangle_\mathrm{max}-\langle\hat{N}_\mathrm{a}\hat{N}_\mathrm{b}\rangle_\mathrm{min}}{\langle\hat{N}_\mathrm{a}\hat{N}_\mathrm{b}\rangle_\mathrm{max}+\langle\hat{N}_\mathrm{a}\hat{N}_\mathrm{b}\rangle_\mathrm{min}}
\end{equation}
where $\hat{N}_\mathrm{a}$ and $\hat{N}_\mathrm{b}$ are the photon number operators for the beamsplitter output modes $a$ and $b$, and $\langle\hat{N}_\mathrm{a}\hat{N}_\mathrm{b}\rangle_\mathrm{max}$ and $\langle\hat{N}_\mathrm{a}\hat{N}_\mathrm{b}\rangle_\mathrm{min}$ are the maximum and minimum  measured coincidence probabilities as we scan through the difference in arrival time (\emph{i.e.} the degree of photon distinguishability) between the incident photons. This definition for the visibility is particularly useful since it is invariant under constant scaling of $\langle\hat{N}_\mathrm{a}\hat{N}_\mathrm{b}\rangle$ and can therefore be applied directly to photon count rates, which are experimentally more readily accessible than expectation values.

\begin{figure}[!htb]
\includegraphics[width=0.8\columnwidth]{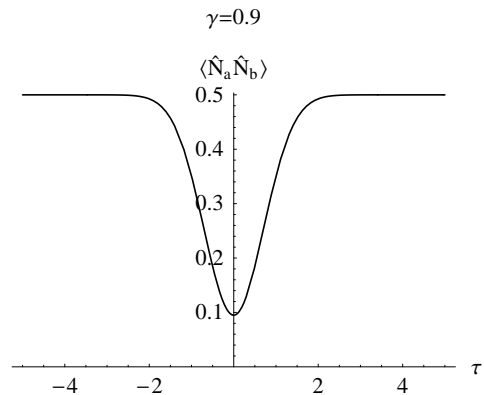}
\caption{HOM dip in the presence of mode-mismatch. The lower bound on $\langle\hat{N}_\mathrm{a}\hat{N}_\mathrm{b}\rangle$ is enforced by mode-mismatch in the non-temporal degrees of freedom.} \label{fig:HOM_dip_mismatch}
\end{figure}

Mode-mismatch has the effect of placing a lower bound on the degree of photon indistinguishability. In other words, even when $\tau=0$ the photons will not be completely indistinguishable as a result of distinguishability in other degrees of freedom (for example spatially). This has the effect of placing a lower bound on $\langle\hat{N}_\mathrm{a}\hat{N}_\mathrm{b}\rangle_\mathrm{min}$, which ideally ought to be 0, resulting in visibilities below unity. This behavior is illustrated in Fig.~\ref{fig:HOM_dip_mismatch}. $\gamma$ is the integral overlap of the photon wave-functions in the non-temporal degrees of freedom and characterizes the degree of mode-mismatch,
\begin{equation}
\gamma=\left|\int_{-\infty}^{\infty}\!\dots\!\int_{-\infty}^{\infty}\psi_\mathrm{A}(\vec\omega_\mathrm{A})\psi_\mathrm{B}^*(\vec\omega_\mathrm{B})\,\mathrm{d}\vec\omega_\mathrm{A}\mathrm{d}\vec\omega_\mathrm{B}\right|^2
\end{equation}
where $\vec\omega_\mathrm{A}$ and $\vec\omega_\mathrm{B}$ are the non-temporal degrees of freedom of photons $A$ and $B$ respectively. $\gamma$ obeys $0\leq\gamma\leq1$, where $\gamma=1$ corresponds to perfect mode-matching and complete photon indistinguishability.

When employing ideal detectors (\emph{i.e} unit efficiency, zero dark-counts, infinite bandwidth and zero resolution), the coincidence rate can be expressed
\begin{equation} \label{eq:HOM_NaNb}
\langle\hat{N}_\mathrm{a}\hat{N}_\mathrm{b}\rangle=\frac{1}{2}-\frac{1}{2}\gamma e^{-\tau^2}
\end{equation}
Upon applying our general detector model we find that $\langle\hat{N}_\mathrm{a}\hat{N}_\mathrm{b}\rangle$ retains this form, up to some constant factor which characterizes the probability of the detectors responding to an incident photon. As discussed previously, HOM visibility is invariant under constant scaling of $\langle\hat{N}_\mathrm{a}\hat{N}_\mathrm{b}\rangle$. Thus, while the rate of photon counts will vary heavily depending on the detector characteristics (specifically the efficiency and bandwidth), the actual HOM visibility is independent of the detectors' efficiency, resolution and bandwidth, provided they are identical. This result is in stark contrast to our observations in the previous section, where detector characteristics were of critical importance.

\section{Conclusion}
We presented a physically motivated, general model for photo-detection. Our models allow for the effects of finite efficiency, resolution and bandwidth, as well as dead-time and dark-counts. Our ideas can be applied to any photonic degree of freedom, including the spectral/temporal, spatial and polarization degrees of freedom. We demonstrated our model by application to two simple, well known situations -- the conditional production of single photons from entangled pairs, and HOM interference. In the former case, detector characteristics were found to be of fundamental importance and have an enormous impact on the operation of the scheme. This is in contrast to the later case, whose operation is independent of the detectors spectral characteristics. This highlights the fact that the effects of photo-detector characteristics are highly application dependent, but, in general, of significant importance. Having a general model for photo-detection is important to modern quantum optics applications such as optical quantum information processing.

\begin{acknowledgments}
This work was supported by the Australian Research Council and the QLD State Government.
\end{acknowledgments}

\bibliography{paper.bib}

\end{document}